\documentclass[preprint]{vgtc}               

\ifpdf
  \pdfoutput=1\relax                   
  \usepackage{graphicx}                
  \DeclareGraphicsExtensions{.pdf,.png,.jpg,.jpeg} 
\else
  \usepackage{graphicx}                
  \DeclareGraphicsExtensions{.eps}     
\fi%

\graphicspath{{figures/}{pictures/}{images/}{./}} 

\usepackage{microtype}                 
\PassOptionsToPackage{warn}{textcomp}  
\usepackage{textcomp}                  
\usepackage{mathptmx}                  
\usepackage{times}                     
\usepackage{cite}                      
\usepackage{tabu}                      
\usepackage{booktabs}                  

\onlineid{1154}

\vgtccategory{Research}

\vgtcinsertpkg

\usepackage{algorithm}
\usepackage{algpseudocode}
\usepackage{amsfonts}
\usepackage{subcaption}

\newcommand{\reals}{\mathbb{R}}
\newcommand{\setdef}[2]{\left\{ #1 \,\middle|\, #2 \right\}}

\usepackage{pdfpages}

\usepackage{tikz}

\definecolor{lime}{HTML}{A6CE39}
\DeclareRobustCommand{\orcidicon}{%
    \begin{tikzpicture}
    \draw[lime, fill=lime] (0,0) 
    circle [radius=0.16] 
    node[white] {{\fontfamily{qag}\selectfont \tiny \textbf{ID}}};
    \draw[white, fill=white] (-0.065,0.1) 
    circle [radius=0.007];
    \end{tikzpicture}
    \hspace{-2mm}
}
\newcommand{\orcid}[1]{\href{https://orcid.org/#1}{\orcidicon}}

\title{Fast Fiber Line Extraction for 2D Bivariate Scalar Fields}

\author{Felix Raith\orcid{0000-0002-3505-6130}\thanks{e-mail: raith@informatik.uni-leipzig.de}\\ %
        \scriptsize Leipzig University %
\and Baldwin Nsonga\orcid{0000-0002-0651-952X}\thanks{e-mail: nsonga@informatik.uni-leipzig.de}\\ %
        \scriptsize Leipzig University %
\and Gerik Scheuermann\orcid{0000-0001-5200-8870}\thanks{e-mail: scheuer@informatik.uni-leipzig.de}\\ %
        \scriptsize Leipzig University %
\and Christian Heine\orcid{0000-0001-7067-5650}\thanks{e-mail: heine@informatik.uni-leipzig.de}\\ %
        \scriptsize Leipzig University %
}

\teaser{
  \centering
  \includegraphics[width=\linewidth]{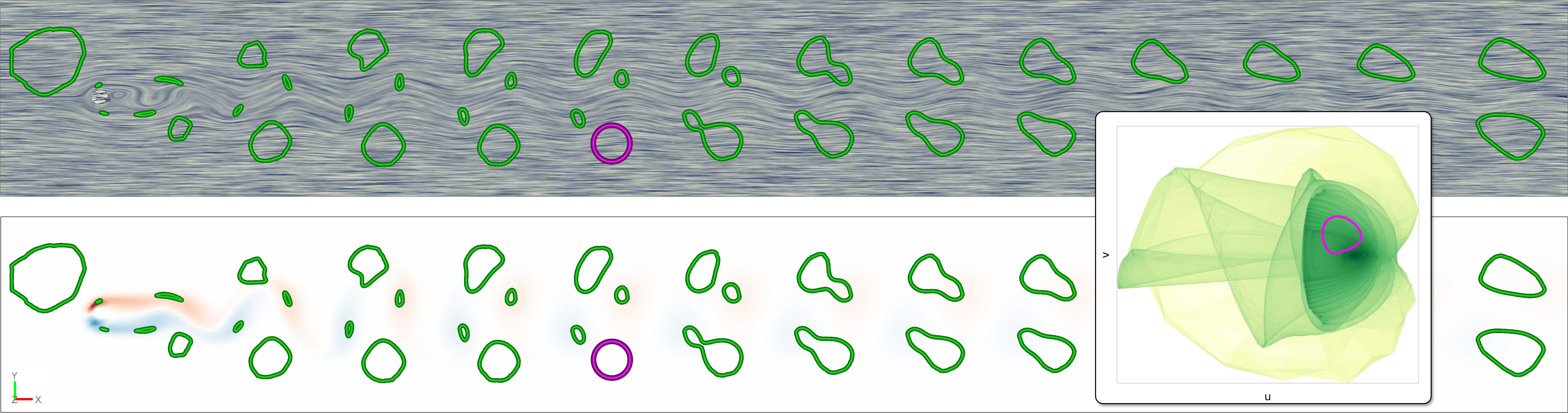}
  \caption{A 2D vector field of a K\'{a}rm\'{a}n vortex street, with a purple curve marking a part of the domain and the preimage of the curve's image in green.
  The inset shows a continuous scatterplot~\cite{Bachthaler:2008:continuous_scatterplots},
  with the curve's image in purple.
  }
  \label{fig:teaser}
}

\abstract{Extracting level sets from scalar data is a fundamental operation in visualization with many applications.
Recently, the concept of level set extraction has been extended to bivariate scalar fields.
Prior work on vector field equivalence, wherein an analyst marks a region in the domain and is shown other regions in the domain with similar vector values, pointed out the need to make this extraction operation fast, so that analysts can work interactively.
To date, the fast extraction of level sets from bivariate scalar fields has not been researched as extensively as for the univariate case.
In this paper, we present a novel algorithm that extracts fiber lines, i.e., the preimages of so called control polygons (FSCP), for bivariate 2D data by joint traversal of bounding volume hierarchies for both grid and FSCP elements.
We performed an extensive evaluation, comparing our method to a two-dimensional adaptation of the method proposed by Klacansky et al., as well as to the naive approach for fiber line extraction. 
The evaluation incorporates a vast array of configurations in several datasets. 
We found that our method provides a speedup of several orders of magnitudes compared to the naive algorithm and requires two thirds of the computation time compared to Klacansky et al. adapted for 2D.%
} 

\CCScatlist{
  Visualization, scalar fields, bivariate data, fibers, preimage extraction algorithm, dual bounding volume hierarchy traversal.
  }

\begin{document}


\firstsection{Introduction \& Background}

\maketitle
Level set methods such as isosurfaces and isolines are fundamental analytic tools in the study of scalar field data, because they often mark useful separations in the data.
Much research effort in visualization was put into making their computation fast, as is evidenced in the Visualization Handbook~\cite{vishandbook2005}, especially to allow interactive analysis.
In many real-world domains such as mechanical engineering, fluid mechanics, and environmental sciences, it is beneficial to investigate the multiple scalar quantities jointly.
However, visual analysis of such multivariate data is generally challenging \cite{Hansen2014}.
Carr et al.~\cite{carr2015fiber} presented fiber surfaces as a conceptual extension of isosurfaces to bivariate data, but to date there has only been limited research effort to make their extraction fast.

Formally, a \emph{level set} of a function $f: D \rightarrow \reals$, $D \subseteq \reals^d$ to a value $v \in \reals$ is the preimage of $v$ under $f$: $f^{-1}(v)=\setdef{x \in D}{f(x) = v}$.
Analogously, a \emph{fiber} of a bivariate function $f: \mathbb{R}^d \rightarrow \reals^2$ to a combination of values $(u,v)\in \reals^2$ is $(u,v)$'s preimage with respect to $f$: $f^{-1}(u,v)=\setdef{x \in D}{f(x) = (u,v)}$, i.e., all points in the domain where $f$ takes on a particular value combination.
While level sets are usually $(d-1)$-dimensional objects, e.g., surfaces in 3D, fibers of bivariate functions are typically $(d-2)$-dimensional objects, e.g., lines in 3D.
Carr et al. \cite{carr2015fiber} introduced \emph{fiber surface control polygons} (FSCP) $P$ approximating a closed curve in the range of $f$ (also called $f$'s \emph{codomain}), defined \emph{fiber surfaces} as the preimage of $P$ under $f$: $f^{-1}(P)=\setdef{x \in D}{f(x) \in P}$, and showed that this typically gives a surface in 3D again.
For users to specify and manipulate the control polygon interactively, the fiber extraction has to be fast enough.

The algorithm by Carr et al. \cite{carr2015fiber} used distance fields to a closed non-self-intersecting control polygon, but was not exact.
Klacansky et al.~\cite{klacansky2017fast} presented an algorithm for both fast and exact extraction of fiber surfaces for bivariate 3D scalar fields using piecewise-linear interpolation, i.e., the bivariate function is defined on a tetrahedral grid and each grid cell uses linear interpolation.
For each edge $e$ of the control polygon and each cell $c$ of the grid, $f$'s preimage of $e$ restricted to $c$ can be computed exactly: (a) $e$ is elongated to an infinite line $l$, (b) the signed distance to $l$ is computed for each vertex of $c$, (c) marching tetrahedra is applied on $c$ with the signed distance field and (d) the result is clipped to the original range of $e$.
The fiber surface is then the union of the preimages for all combinations of $m$ edges with $n$ cells, giving a run-time in $O(m\cdot n)$.

The cell/edge preimage computation involves many arithmetic operations and tests, making extraction using this naive algorithm prohibitively slow for interactive applications.
Additionally, only few cell/edge combinations have a nonempty preimage, i.e. contribute to the fiber surface.
For these two reasons, Klacansky et al.~\cite{klacansky2017fast}'s method computes a \emph{bounding volume hierarchy} (BVH) using \emph{axis-aligned bounding boxes} (AABBs) of the cell's images and queried it for each polygon edge using an adaptation of well known ray casting BVH traversal techniques.
There is an overhead for the construction of the BVH, but this has to be performed only once if all interaction affects the control polygon only.
Also, the result of the bounding volume hierarchy traversal can yield \emph{false positives}, i.e., cell/edge combinations that do not intersect even though the bounding box of the cell intersects the edge.
In practice, the algorithm's run-time grows nearly logarithmic with the number of cells, but it is still linear in the number of control polygon's edges.
Klacansky et al.~\cite{klacansky2017fast} considered control polygons only up to 11 edges and reported an average speed-up of roughly one order of magnitude compared to the naive algorithm.
In many applications of fiber surfaces (see \autoref{fig:teaser} and \autoref{sec:relatedwork}), the control polygons can comprise many more edges and raises the question for algorithms that scale sublinearly with the number control polygon edges.

Our work aims to further speed up the extraction of preimages for bivariate data when the number of control polygon edges is large.
In this paper we focus on 2D domains, like Zheng et al.~\cite{zheng2019visualization}, who gave use cases for complex control polygons.
We will refer to the extracted features as \emph{fiber lines}.
Central to our approach is the use of two BVHs, one for the cells and one for the control polygon edges, and traversing both BVHs jointly.
This \emph{Dual-BVH traversal (DBt)} common practice in computer graphics and robotics for collision detection and can be described as a recursive algorithm:
Given a pair of tree nodes $l, r$ from both trees, if their bounding volumes intersect, recurse into either $l$'s children or $r$'s children -- usually based on the size of $l$'s and $r$'s bounding volume.
The algorithm is initialized with the roots of both BVHs and reports all combinations of leaves whose bounding volumes intersect.
For consistency, we use AABBs as bounding volumes for the edge BVH, too, even though, this will most likely produce more false positives compared to Klacansky et al.~\cite{klacansky2017fast}.
On the other hand, the dual traversal reduces the number of intersection tests, each of which use fewer arithmetic operations.
However, it is not immediately clear whether the reduction in intersection test costs is worth the overhead of building the BVH for the control polygon and treatment of additional false positives.
We find that a blend of Dual-BVH and Klacansky combines the best of both methods.
The contributions of this paper are: (1) a novel algorithm for the extraction of fiber lines, combining the approach by Klacansky et al. with \emph{DBt}, and (2) results from an extensive evaluation of the different methods, studying factors such as control polygon size and recursion strategies, in both theoretical and practical test settings.

\section{Related Work}
\label{sec:relatedwork}

An overview of visualization for multivariate data is given by Fuchs and Hauser \cite{Fuchs2009}.
Standard techniques for the depiction of multivariate data include glyphs \cite{chung2014glyph}.
In visual analysis it is typical to select important features interactively, but
there are also automatic approaches, e.g. the work by J\"anicke et al. \cite{janicke2007multifield}.
The importance of feature-based techniques was highlighted in the work of Obermaier et al. \cite{obermaier2014feature} and Carr \cite{carr2014feature}.
Sauber et al. \cite{Sauber2006} used a multifield graph for 3D scalar fields.
Each node corresponds to a correlation field combining at least two fields, aiding the selection which correlation fields to investigate further.
Nagaraj et al. \cite{Nagaraj2011} define a measure of scalar field similarity based on the field's gradients' similarity.

Fiber surfaces as an extension of isosurfaces to multifields were proposed by Carr et al.~\cite{carr2015fiber} based on earlier work by Saeki~\cite{Saeki2004} on singular fibers.
The method determines the preimage of a control polygon in the range of a bivariate function as the isosurface to the zero-level of a field coding the signed distance to the control polygon.
It restricted the control polygon to be closed and non-self-intersecting, and only gave an approximate result, because the signed distance field was stored using piecewise interpolation.
Wu et al.~\cite{Wu2017} presented a volume ray casting algorithm to display fiber surfaces.
A different approach is taken by Jankowai and Hotz~\cite{jankowai2018feature}.
They use feature level sets with traits to define so-called standard isosurfaces, and fiber surfaces are special cases of these feature level sets.
Klacansky et al.~\cite{klacansky2017fast} presented an exact method to compute fiber surfaces for bivariate functions defined on tetrahedral meshes using piecewise linear interpolation, and introduced using spatial search structures for objects in the function's range.

Concerning applications, Sakurai et al.~\cite{sakurai2016} presented a tool to study singular fibers interactively, and in another work, Sakurai et al.~\cite{sakurai2020} generalized the flexible isosurfaces interface~\cite{Carr2010} to fiber surfaces.
Tierny and Carr~\cite{tierny2017} link Jacobi sets~\cite{Edelsbrunner2004} and Reeb spaces~\cite{Edelsbrunner2008} with fiber surfaces by using the Jacobi sets as seeds to determine fiber surface components that subdivide the domain into Reeb space components.
Zheng et al.~\cite{zheng2019visualization} study 2D vector field equivalence, which can be expressed as the preimage of the image of an object in the domain of the function.
They noted that their method was very slow.
Raith et al. \cite{raith2019tensor,raith2020visual} study trivariate functions arising from the study of 3D tensor field invariants, and its application and usefulness were shown in different settings \cite{Blecha2019analysis,raith2021uncertainty}.
Their approach has since been extended to general multivariate data by Blecha et al. \cite{blecha2020}.
Recently, Sharma et al. \cite{sharma_continuous_2023} used fiber surface control polygons seeded from continuous scatter plots to analyze electronic transitions.
All these applications have in common that they usually require the extraction of preimages for control polygons and surfaces with many elements in typically interactive settings.
In the work of Sharma and Natarajan \cite{sharma2022jacobi}, the Jacobi set is used to identify interesting barrel surface components. 
For this purpose, they present an output-sensitive approach for their calculation.

Our work is based on well-established results from computer graphics in the area of collision detection. A survey is given by \cite{sulaiman2012bounding}.

\section{Method}
\label{sec:algorithm}
In this section, we describe our approach of combining \emph{Dual-BVH traversal (DBt)} with Klacansky's approach.
We also address important optimization issues such as the choice of recursion strategies, and explain the importance of the false positive ratio.
In doing so, we show how this can be reduced without incurring a time loss in the corresponding preprocessing step.
The algorithm here consists of two consecutive parts, i.e., a search space reduction using DBt (cf. Algorithm \ref{alg:SpeedUp}) followed by an extraction algorithm.

\subsection{Extraction Algorithm}
We begin by briefly elaborating on the base extraction algorithm, which is a standard implementation for extracting fiber lines adjusted for two-dimensional data.
Given the field $f: D \to \mathbb{R}^2$, with $D \subseteq \mathbb{R}^2$ defined on a triangular mesh mapped to the range of $f$ in which also the FSCP is defined.
The algorithm now computes the coefficients for the transformation of the FSCP edges from the range to the domain $D$.
For this purpose, the line coordinates are first normalized for each edge.
If thereby an edge has a length of $0.0$, it is skipped.
Then we iterate over each triangular cell of the mesh for the current edge and calculate the signed distance between the edge and every vertex of the triangle.
Using the signed distances, we apply marching triangles.
We then calculate the intersection points of the FSCP edge with the triangle edges normalized to the values between 0 and 1.
If both intersection points are on one side of the triangle, the triangle is skipped, otherwise, the line is truncated inside the triangle.
Finally, we map the resulting intersection points to the domain $D$.
A pseudocode implementation for the extraction algorithm can be found in the supplemental material.

\subsection{Acceleration structure}

As discussed earlier, Klacansky et al. \cite{klacansky2017fast} accelerate the algorithm by using a BVH.
Instead of directly applying the fiber surface extraction, they first perform intersection tests between AABB and the FSCP edges. 
Our idea is to replace the intersection tests between lines and AABBs with intersection tests between pairs of AABBs. 
As the corresponding AABBs can be represented as two sets of coordinates spanning the AABBs, an intersection test between two AABBs amounts to a comparison of their coordinates.
In addition to a reduced computational cost, using a BVH for the FSCP promises an efficient search space reduction, also for complex FSCP with many edges. 
When intersecting using Klacansky's approach, a $m$ number of FSCP edges and a BVH for the bivariate field in the range of $f$, each line must be compared to the BVH, resulting in a time-complexity linear in $m$ and in theory logarithmic in $n$, where $n$ is the number of cells.
Utilizing two BVHs, the time-complexity can be reduced to logarithmic in $m$ in theory. 
A downside is that an AABB does not approximate a line well and therefore tests between two AABB may lead to more false positives than tests between an AABB and a line.
To prevent this, once the BVH for the FSCP is in the leaves we use the test between an AABB and a line.
This allows us to benefit from the low-cost calculation of the intersection between two AABBs, and at the same time from the more accurate tests between an AABB and a line.
This provides an overall advantage for complex FSCP, making it more suitable for use in real-time applications.

We denote $T$ as a BVH, which can either be constructed using the FSCP edges or using the triangles in the range as input.
Our acceleration structure is then as follows:
We first construct the two BVH $T_1$ and $T_2$ using the triangles and FSCP edges, respectively.

First, we test if the AABB in the roots $T_1$ and $T_2$ overlap at all. 
If this is not the case, the algorithm is aborted. 
Otherwise, we subdivide $T_1$ and check if it is child $c$ and $T_2$ are leaves.
If this is the case then a cell-line pair is formed and stored accordingly in the list $R$.
If this is not the case, it is checked whether all children $c$ intersect with $T_2$ using the intersection strategy.
This intersection strategy is the most important part of the algorithm and decides if $c$ or $T_2$ is a leaf node and at the same time contains the FSCP edges, that we use the tests between an AABB and a line.
Otherwise, we use the test between two AABB for the intersection test. 
If they do not intersect, the corresponding $c$ is skipped.
Otherwise, the recursion strategy is used to check whether $c$ and $T_2$ must be interchanged and the algorithm is called recursively accordingly.
The algorithm returns the list $R$ at the end, which contains all possible cell line pairs.\begin{algorithm}
\caption{\textbf{Speedup} with BVH Algorithm}\label{alg:SpeedUp}
\begin{algorithmic}[1]
\Require BVH from dataset in codomain (triangles) $T_1$, BVH from control polygon in codomain (lines) $T_2$, return value $R$
\State set $n_1$ = rootOf($T_1$), $n_2$ = rootOf($T_2$)
\Function {traverse} {$n_1$, $n_2$, $R$}:
\ForAll{children $c$ of $n_1$}
    \If{isLeaf($c$) \textbf{and} isLeaf($n_2$)}
        \State \textbf{add} combination from cells in $c$ \& $n_2$ to $R$
    \ElsIf{\textbf{intersectionStrategy}($c$, $n_2$)}
        \If{ \textbf{recursionStrategy} }
            \State TRAVERSE ($c$, $n_2$, $R$)
        \Else
            \State TRAVERSE ($n_2$, $c$, $R$)
        \EndIf
    \EndIf
\EndFor
\EndFunction
\end{algorithmic}
\end{algorithm}
\vspace{-0.5em}

\subsection{Recursion Strategies}
When searching through two BVHs, it is important to choose a reasonable strategy for selecting which BVH to reduce at every step to optimize search time, especially when BVH have different heights.
A common approach is to select the tree with the greater height to reduce at each step, as this generally results in a faster overall search. 
However, it is also important to consider other factors such as the size and structure of the BVHs, as well as the search algorithm used, to determine the most effective reduction strategy.
In addition, it may be beneficial to use heuristics or other techniques to guide the search and improve its efficiency.
Ultimately, the choice of reduction strategy depends on the specific requirements and constraints of the search problem at hand.
When investigating different swap strategies, we found the recursion strategy \textit{area} to be most effective. This strategy is defined as the area of the current bounding volume processed by the algorithm. 
A detailed analysis of the investigated recursion strategies can be found in the supplemental material.

\section{Results}
We demonstrate our algorithm and perform a comprehensive evaluation of three use cases.
These are (I) the theoretical case, (II) the practical case, and (III) the field equivalence case.
First, we describe the data sets and FSCP used.
We then use the cases to compare our method with the comparison algorithms, the naive approach, DBt, and the Klacansky algorithm.
For Klacansky, we used $8$ elements per leaf, as this was reported as the best trade-off by Klacansky et al.\cite{klacansky2017fast}.
For DBt and the blend between DBt and Klacansky we use $1$ elements per leaf since we found this to give the best trade-off between search time and extraction time.
To compare the search structures in isolation, we use the same extraction algorithm (see \autoref{sec:algorithm}) for fiber lines with all search structures.
All tests were performed on a workstation with an Intel Xeon E5-2630 v3 running at 2.40 GHz, 32 GB RAM, and an NVIDIA GeForce GTX 980. 
All algorithms are implemented using C/C++ and are not parallelized.

\subsection{Datasets \& Control Polygons}
\begin{figure}[tbp]
  \centering
  \begin{subfigure}[b]{0.24\linewidth}
  	\centering
  	\includegraphics[width=\textwidth]{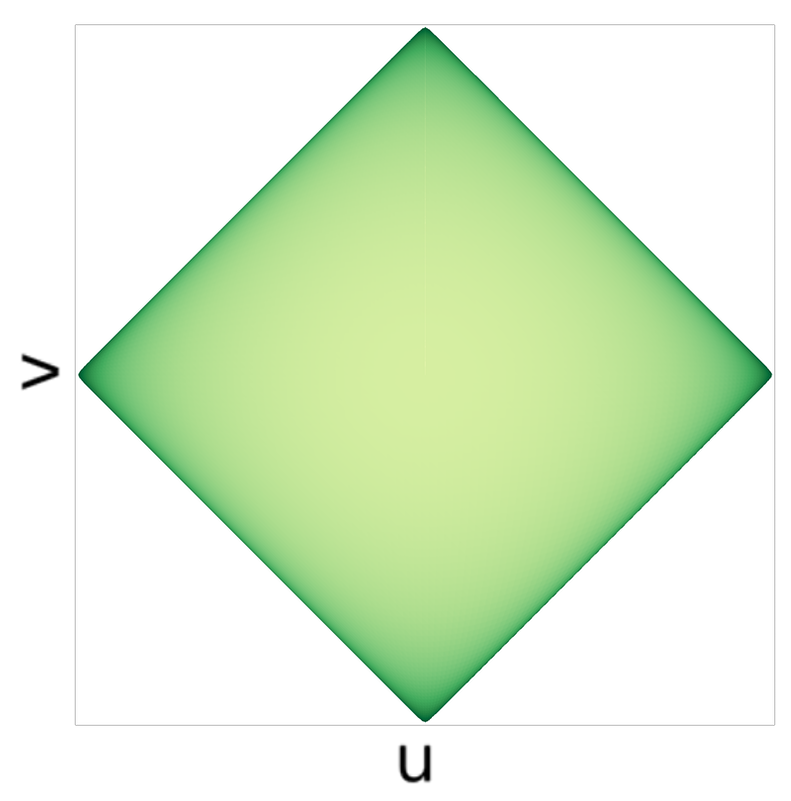}
  	\caption{Double Gyre}
  	\label{fig:datasets_subfigs_a}
  \end{subfigure}%
  \begin{subfigure}[b]{0.24\linewidth}
  	\centering
  	\includegraphics[width=\textwidth]{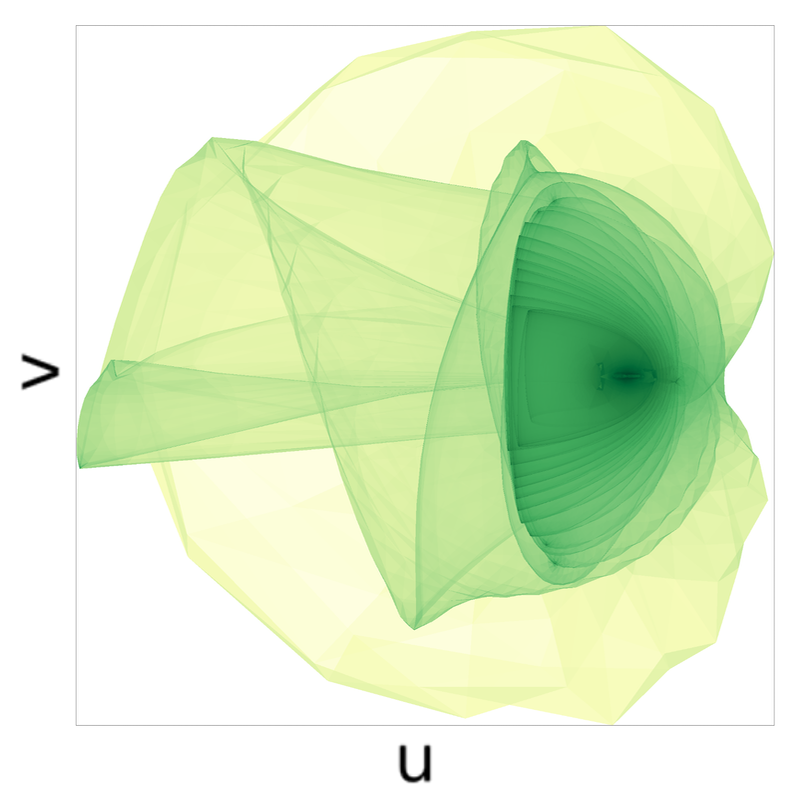}
  	\caption{Vortex Street}
  	\label{fig:datasets_subfigs_b}
  \end{subfigure}%
    \begin{subfigure}[b]{0.24\linewidth}
  	\centering
  	\includegraphics[width=\textwidth]{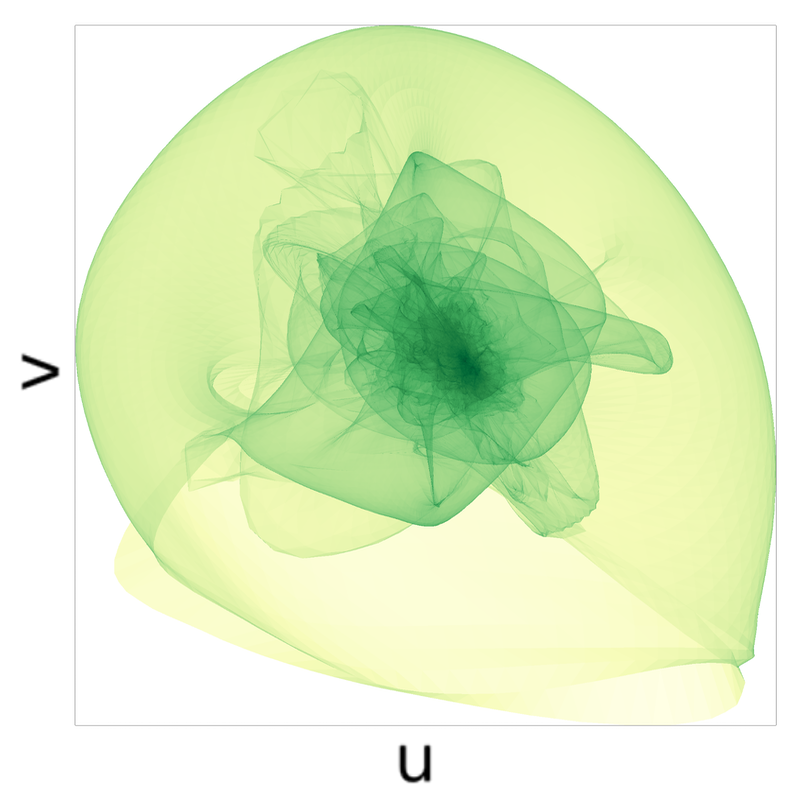}
  	\caption{Red Sea}
  	\label{fig:datasets_subfigs_c}
  \end{subfigure}%
    \begin{subfigure}[b]{0.24\linewidth}
  	\centering
  	\includegraphics[width=\textwidth]{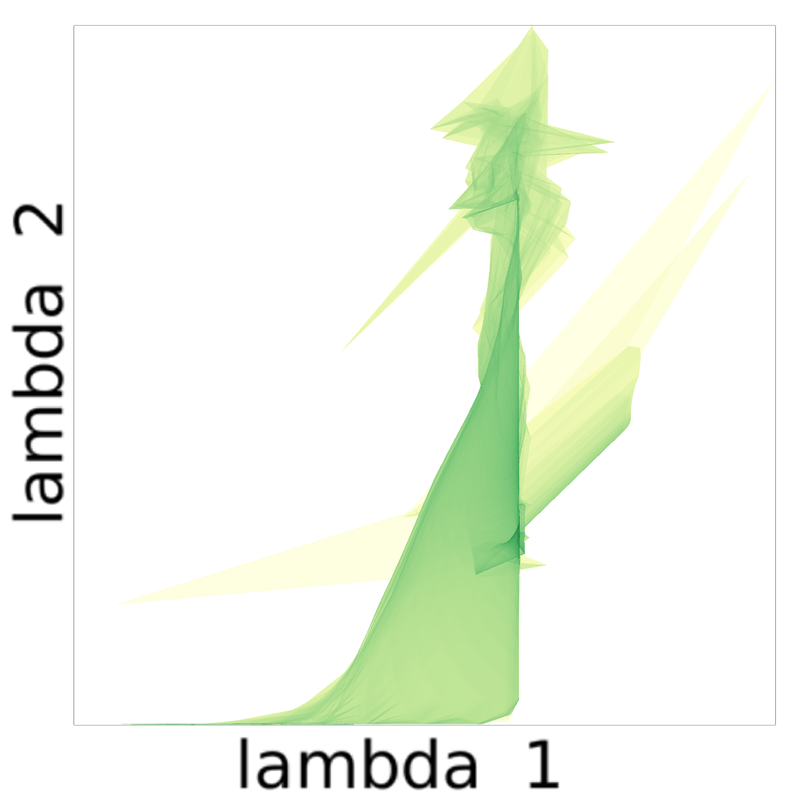}
  	\caption{Tensile Bar}
  	\label{fig:datasets_subfigs_d}
  \end{subfigure}%
  
    \begin{subfigure}[b]{0.24\linewidth}
  	\centering
  	\includegraphics[width=\textwidth]{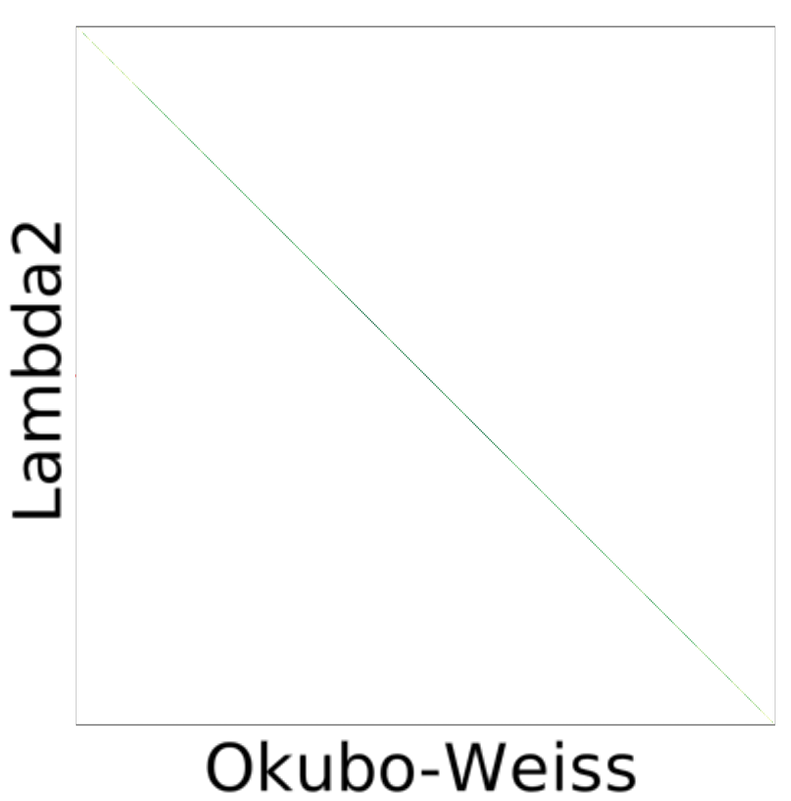}
  	\caption{Double Gyre}
  	\label{fig:datasets_subfigs_e}
  \end{subfigure}%
    \begin{subfigure}[b]{0.24\linewidth}
  	\centering
  	\includegraphics[width=\textwidth]{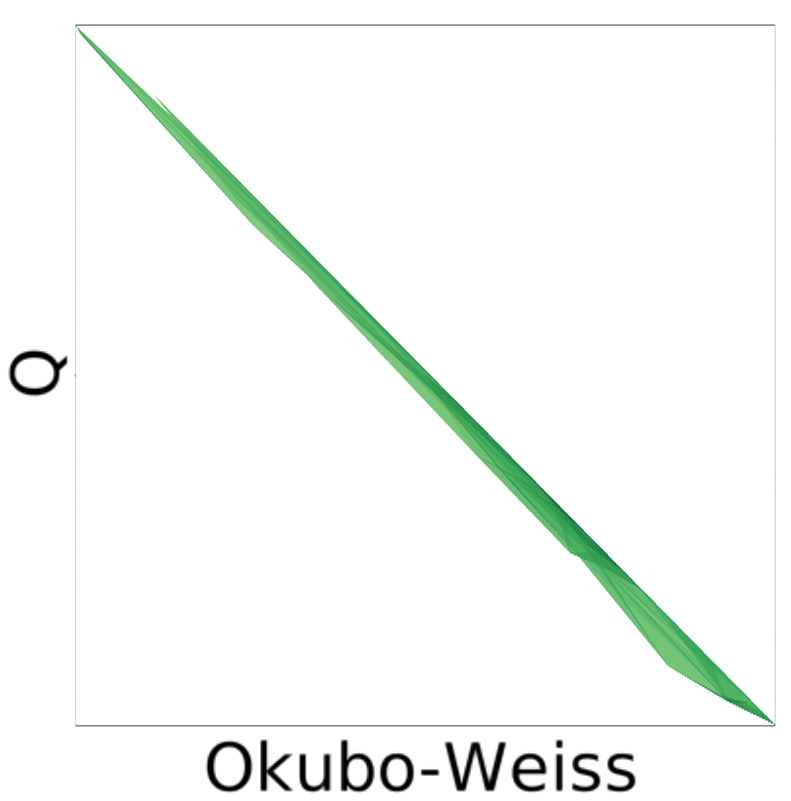}
  	\caption{Vortex Street}
  	\label{fig:datasets_subfigs_f}
  \end{subfigure}%
    \begin{subfigure}[b]{0.24\linewidth}
  	\centering
  	\includegraphics[width=\textwidth]{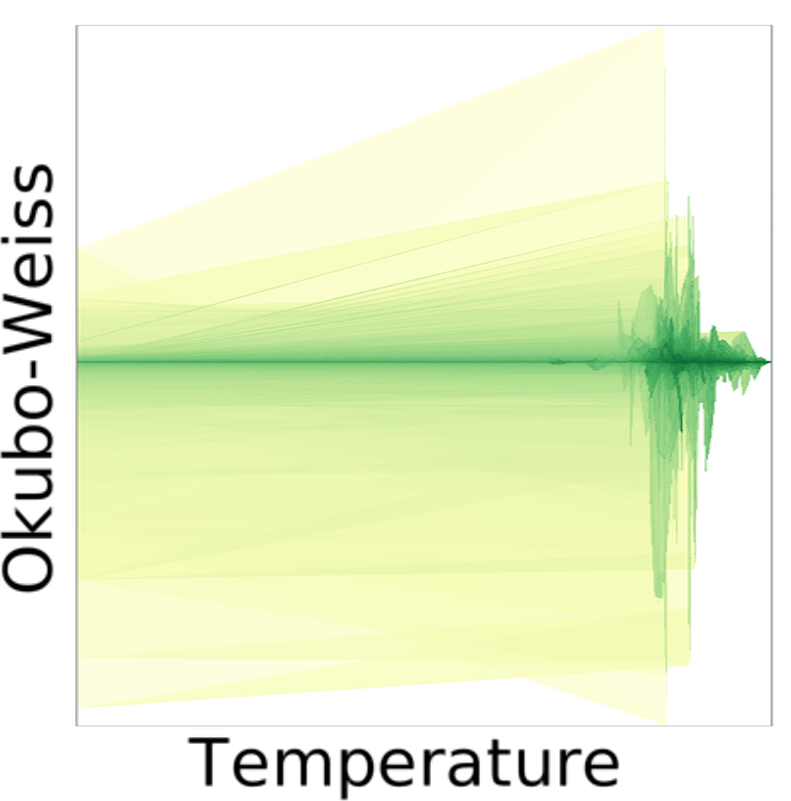}
  	\caption{Red Sea}
  	\label{fig:datasets_subfigs_g}
  \end{subfigure}%
  \begin{subfigure}[b]{0.24\linewidth}
  	\centering
  	\includegraphics[width=\textwidth]{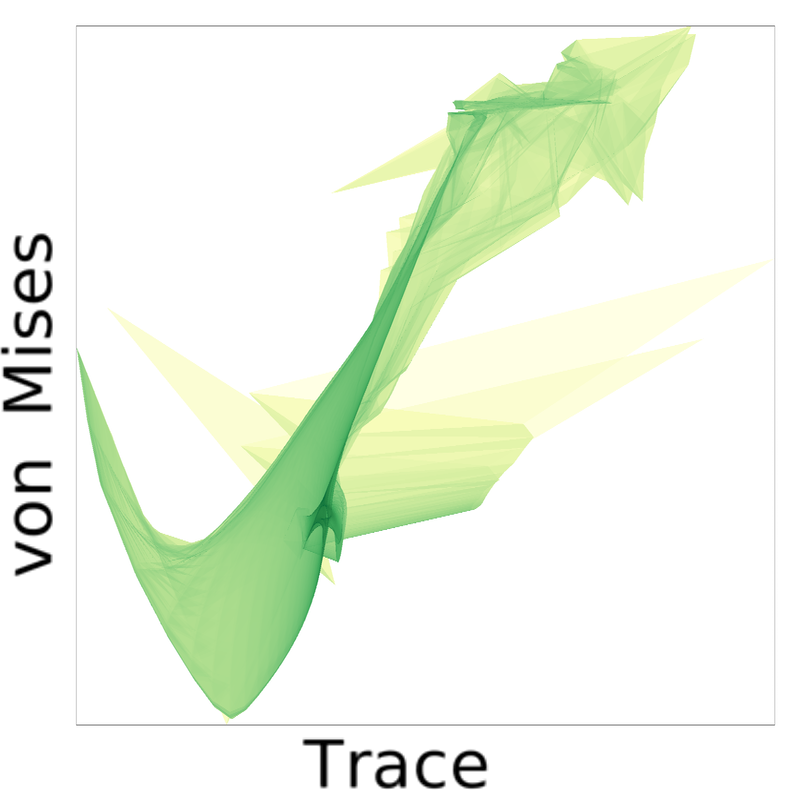}
  	\caption{Tensile Bar}
  	\label{fig:datasets_subfigs_h}
  \end{subfigure}%
  \caption{Continuous scatterplot for each dataset's range.}
  \label{fig:datasets}
\end{figure}

In this work, we use four well-known freely available datasets from the literature, whose continuous scatterplots \cite{Bachthaler:2008:continuous_scatterplots} are represented in \autoref{fig:datasets}.
One synthetic dataset, the Double Gyre \cite{Shadden05} ($64770$ cells, \autoref{fig:datasets_subfigs_a},e) and three real datasets, the K\'{a}rm\'{a}n Vortex Street from the Topology ToolKit \cite{ttk, ttk19} data package ($65536$ cells, \autoref{fig:datasets_subfigs_b},f), a common flow pattern, the Red Sea dataset from the 2020 SciVis contest \cite{zhan2014eddies, zhan2019three, hoteit2018data} ($498002$ cells, \autoref{fig:datasets_subfigs_c},g), and the Tensile Bar \cite{zobel2017visualizing, zobel2018extremal} ($9000$ cells, \autoref{fig:datasets_subfigs_d},h) from mechanics.

Furthermore, we used nine FSCP for the tests, for marking relevant regions in the codomain.
The FSCPs range from 3 to 2997 edges.
Note that Klacansky et al. only considered control polygons comprising up to 11 edges.

\begin{figure}[tb]
  \centering 
  \includegraphics[width=\columnwidth]{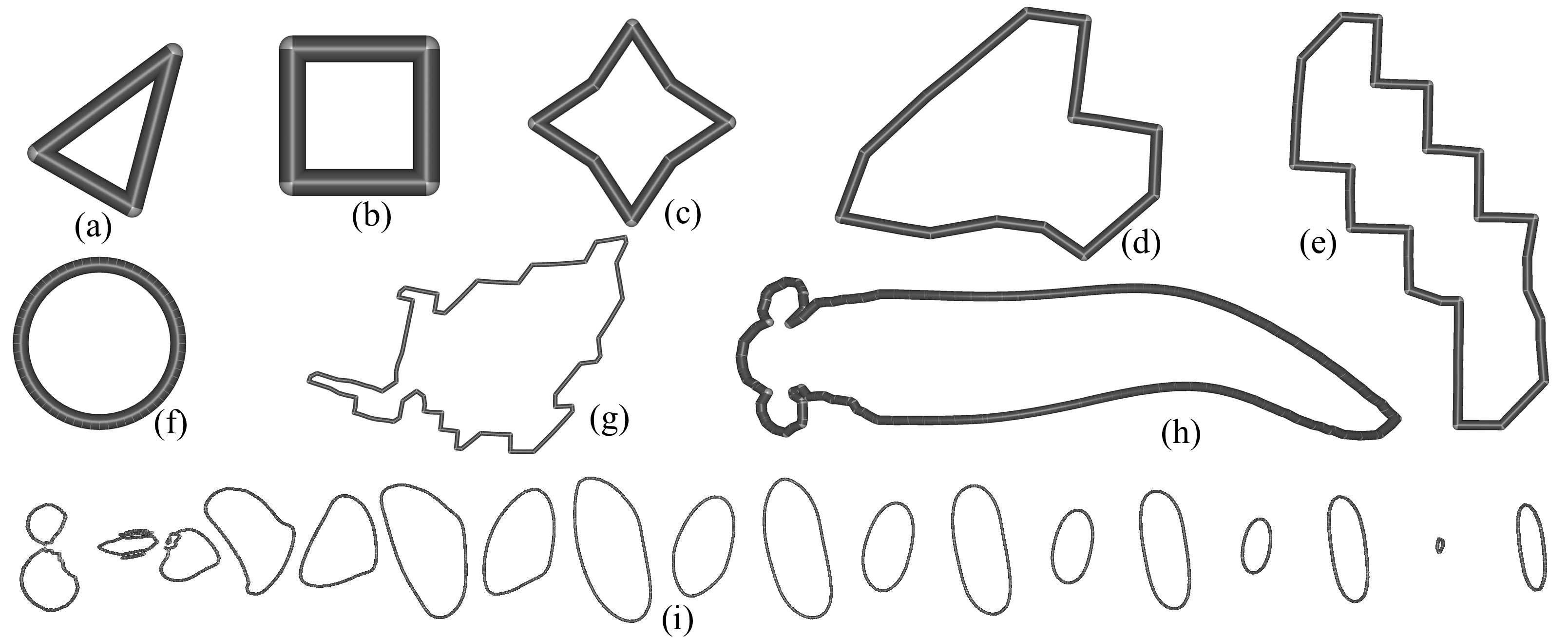}
  \caption{%
  	Visualization of the 9 control polygons in the codomian. Number of edges: (a)=3, (b)=4, (c)=8, (d)=16, (e)=38, (f)=60, (g)=126, (h)=232, and (i)=2997.
  }
  \label{fig:controlPolygons}
\end{figure}
\vspace{-0.5em}

\subsection{Evaluation in the Theoretical Case (I)} 
First, we compare in detail the total time and the quantities that actively affect it, and compare them with the comparison algorithms in case (I). 
We use all the datasets and intersect them with the prefabricated FSCP from \autoref{fig:controlPolygons}.
We position the center of the FSCP at $205$ random positions in the range of AABB of the corresponding dataset codomains given in \autoref{fig:datasets}a,b,c,d.
\begin{table}[tb]
    \caption{
    Evaluation of case (I) with total time consisting of time for creation of additional BVH if available, intersection time, and extraction time in ms, number of intersection tests (NiT), and ratio true positives vs. all positives (TPaP). All values are averages, and bold are the best values.}
    \label{tab:resultsTheoCase}
    \scriptsize%
    \centering%
    \begin{tabu}{*{5}{c}}
        \toprule
        Methode & Dataset & NiT & TPaP & Total Time \\
        \midrule
        Naive& Double Gyre & -- & 0.11\% & 1048.36 \\
        approach & K\'{a}rm\'{a}n V. S. & -- & 0.15\% & 1059.19 \\
        & Red Sea & -- & 0.04\% & 7997.22 \\
        & Tensile Bar & -- & 0.25\% & 154.68 \\
        \midrule
        Klacansky & Double Gyre & \textbf{7160.60} & 50.04\% & 0.87 \\
        BVH & K\'{a}rm\'{a}n V. S. & 43856.45 & 17.95\% & 3.31 \\
        8 elems. & Red Sea & 45205.72 & 23.42\% & 3.96 \\
        per leaf & Tensile Bar & 9271.62 & 24.97\% & 0.99 \\
        \midrule
        DBt & Double Gyre & 16626.74 & 47.94\% & 1.99 \\
        1 elem. & K\'{a}rm\'{a}n V. S. & 35094.51 & 40.06\% & 3.92 \\
        per leaf & Red Sea & 115691.46 & 42.68\% & 5.01 \\
        & Tensile Bar & 17333.14 & 40.38\% & 1.22 \\
        \midrule
        Our & Double Gyre & 9363.42 & \textbf{75.06\%} & \textbf{0.80} \\
        Algorithm & K\'{a}rm\'{a}n V. S. & \textbf{26937.60} & \textbf{54.45\%} & \textbf{1.73} \\
        1 elem. & Red Sea & \textbf{40804.39} & \textbf{61.73\%} & \textbf{2.29} \\
        per leaf & Tensile Bar & \textbf{7765.37} & \textbf{51.08\%} & \textbf{0.65} \\
        \bottomrule
    \end{tabu}%
\vspace{-1em}
\end{table}
The \autoref{tab:resultsTheoCase} shows the evaluation for the total time across all $184,500$ test cases.
Mean values over all test cases are given for the values.
It can be seen, that our algorithm is faster on average over all datasets compared to the other methods and has less search overhead.
For example, our algorithm achieves a speedup of $2$ orders of magnitude over the naive approach, requiring two-thirds of the total time of Klacansky, and is more than twice as fast as the DBt.
The details also show that there is a significant difference between datasets.
For example, the difference in the synthetic dataset is smaller than in the real datasets.
This may be due to the codomain, which is very regular in the analytic dataset and contains hardly any overlapping cells, as can also be seen in the scatterplot of \autoref{fig:datasets_subfigs_a}.

\subsection{Evaluation in the Practical Case (II)}
\begin{figure}[tb]
  \centering 
  \includegraphics[width=\columnwidth]{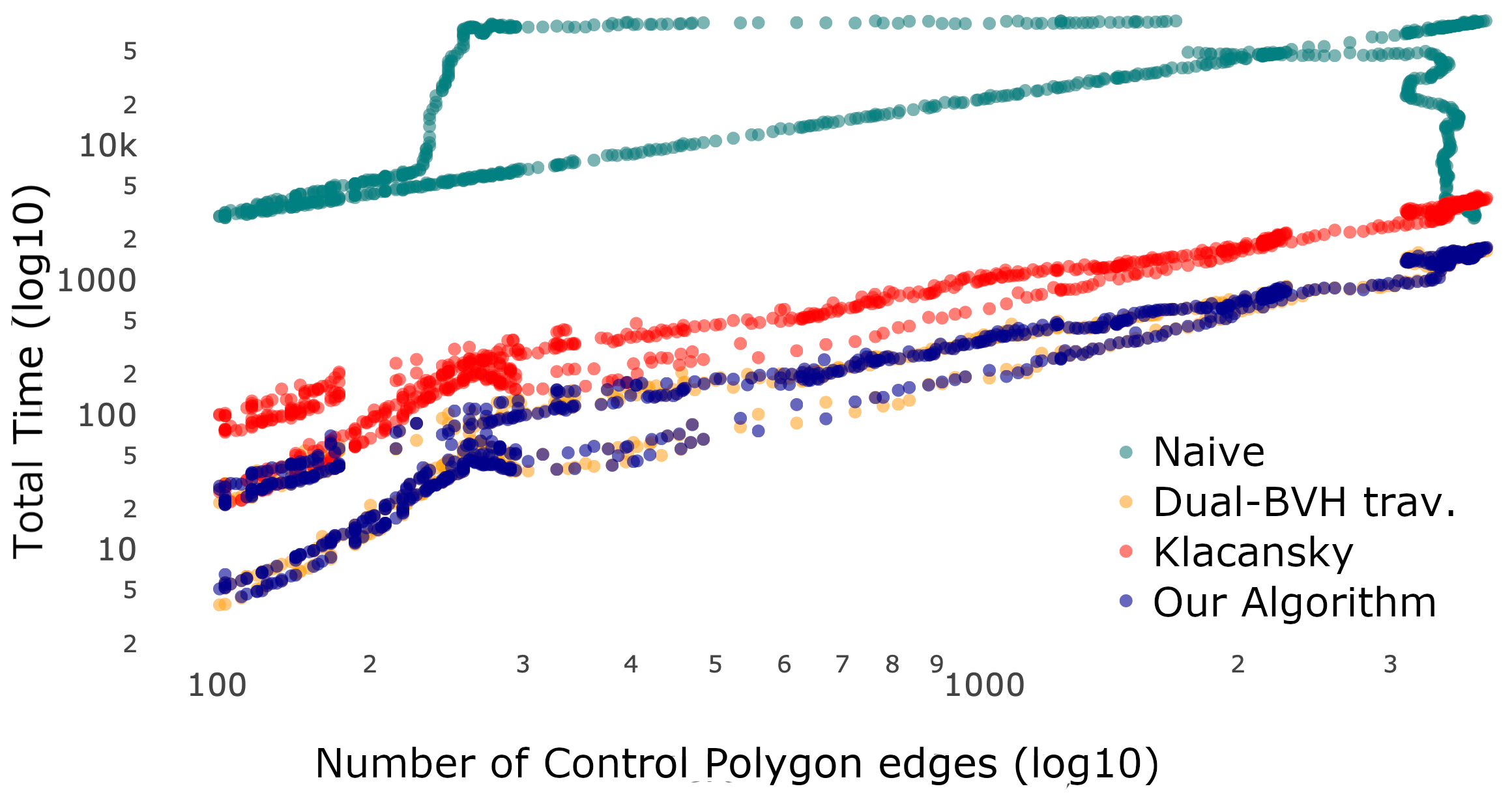}
  \caption{%
  	Plot the dataset Red Sea in the case (II).
  }
  \label{fig:plotPracCase2}
\end{figure}

During the automatic generation of FSCP, a high number of edges are frequently generated, which always lie completely in the codomain of the object.
Therefore, individual edges cannot be excluded at an early stage.
To investigate this for practical use cases, we consider the case (II) and compare the total time.
Case (II) automatically generates a FSCP for the algorithm, in which an isoline is created for a given iso value and projected from the domain into the codomain.
The dataset codomains are shown in \autoref{fig:datasets}e,f,g,h.
We create $1001$ FSCP per dataset by corresponding isovalues.
This gives a total number of $4004$ tests for case (II).
Compared to the naive approach and Klacansky, our algorithm requires less total time (see \autoref{fig:plotPracCase2}).
In particular, for the naive approach, we see that our algorithms are on average $1.67$ orders of magnitude faster and in the best case even $1.76$ orders of magnitude faster, which is even better than in case (I).
Compared to Klacansky, our algorithm is $2.2$ times faster.
Only DBt performs significantly better than in Case (I) and is about as fast.
\vspace{-0.5em}

\subsection{Evaluation in the Field Equivalence Case (III)}
Case (III) is borrowed from the work of Zheng et al.~\cite{zheng2019visualization} in which field equivalence was calculated.
However, as they admitted, their calculations were very slow.
Our algorithm can reproduce the field equivalences by running two algorithms in sequence.
We examine the overall time, i.e., the total time of the first and second algorithms, to determine whether we are also faster than the comparison algorithms.
To mark the field equivalence, we use the FSCP from \autoref{fig:controlPolygons}f and move it over the domain.
Its fiber line is used as the new FSCP, to represent the field equivalence as a fiber line, as shown in \autoref{fig:teaser}.
For our tests, we use the codomain as shown in \autoref{fig:datasets}e,f,g,h.
For replication, we positioned the control polygon at $101$ positions per dataset using the codomain from \autoref{fig:datasets}e,f,g,h, to compute field equivalences, resulting in $404$ test cases.
This results in a mean overall time of $40.23$\,ms, which allows for interactive analysis.
When considering the K\'{a}rm\'{a}n vortex street, which Zheng et al. also considered, we require a mean overall time of $37.99$\,ms, indicating that our algorithm is faster than that of Zheng et al.~\cite{zheng2019visualization}.
We have found that our algorithm is on average $2.2$ orders of magnitude faster than the naive approach, we are $1.75$ times faster than Klacansky, and we are $1.19$ times faster than the DBt.

\section{Conclusion \& Future Work}

In this paper, we have presented a new algorithm that speeds up the extraction of fiber lines for bivariate 2D scalar fields.
We have performed an extensive run-time comparison and shown that our algorithm is several orders of magnitude faster than the naive approach and also requires less total time than the adapted algorithm of Klacansky and DBt.
To this end, we compared the algorithms in three cases. 
We optimized and evaluated individual components of the algorithm in case (I) and compared their final results with naive, DBt, and Klacansky.
We found, that our approach is particularly well suited for FSCP with many edges.
In case (II), we investigated the relationship between the total time and the number of edges. 
In case (III), we used our algorithm to extract field equivalents and similarly this is faster than the naive algorithm and Klacansky.

Our research has shown that the total time is mostly dominated by the extraction time.
Further improvements are thus to be expected from reducing false positive cell/edge combinations.
One could also extend the framework by Zheng et al.~\cite{zheng2019visualization} to use our method.
The results of our work suggest that it would be significantly faster.
Moreover, our results indicate a higher benefit of using our approach over a 2D version of Klacansky et al.~\cite{klacansky2017fast} when the number of edges in the FSCP is large.
In future work, we will study the performance of our algorithm adapted to 3D domains, because most applications focus on this case.










\acknowledgments{This work was funded by the Deutsche Forschungsgemeinschaft (DFG, German
Research Foundation) - SCHE 663/17-1.}

\bibliographystyle{abbrv-doi}

\bibliography{template}






\section*{Supplementary material}

# Fast Fiber Line Extraction for 2D Bivariate Scalar Fields (Supplement)

Felix Raith *    Baldwin Nsonga †    Gerik Scheuermann ‡    Christian Heine §

Leipzig University    Leipzig University    Leipzig University    Leipzig University

## 1 METHOD

Here the extraction algorithm from the main paper is given as pseudocode.

---

**Algorithm 1** Base Fiber Line Extraction Algorithm

---

**Require:** grid $G$, control polygon $CP$
1: **for all** edges $e \in CP$ **do**
2:     **compute** edge length $d$
3:     **if** $d = 0$ **then**
4:         **continue**
5:     **end if**
6:     **extend** $e$ to line $l$
7:     **for all** triangle $t \in G$ **do**
8:         **compute** signed distance $d_i$ to $l$ for each triangle point
9:         **set** case $c$ to 0
10:        **compute** $c$ for all $d_i$
11:        **if** $c = 0$ **then**
12:            **continue**
13:        **else if** $isSwitch(case)$ **then**
14:            **switch** triangle points
15:        **end if**
16:        **compute** intersections of line with triangle edges
17:        **compute** value $t_1$ and $t_2$ at intersections
18:        **if** $\lfloor t_1 \rfloor \cdot \lfloor t_2 \rfloor > 0$ **then**
19:            **continue**
20:        **end if**
21:        **clip** line to interior of triangle
22:        **transform** to domain coordinates
23:        **store** fiber line
24:     **end for**
25: **end for**

---

## 2 RESULTS

### 2.1 Datasets

In this section the used datasets are described in detail.

**Double Gyre** As a first dataset, the Double Gyre by Shadden et al. [6], which can be found on the website of ETH Zurich. In this analytical dataset, two oppositely rotating vortices are simulated. The data is available as a time-dependent vector field with 512 time steps and contains the two velocity components $u$ and $v$. The dataset has a resolution of $256x128$ and the unstructured grid consists of a total of 64770 triangular cells. We use the time step 0 for our investigation.

*e-mail: raith@informatik.uni-leipzig.de
†e-mail: nsonga@informatik.uni-leipzig.de
‡e-mail: scheuer@informatik.uni-leipzig.de
§e-mail: heine@informatik.uni-leipzig.de

**Kármán Vortex Street** Second, we use a numerical simulation of the Kármán Vortex Street by Popinet [4], which can also be found on the Topology ToolKit [3,7] website in the data package [8]. This dataset is a well-known two-dimensional dataset from flow visualization, which shows how a fluid flows past a cylinder in a flow channel. This creates vortices, which are classically visualized. The simulation was created using the Gerris flow solver [5] and contains a vector field and the two velocity components $u$ and $v$. The unstructured grid of the dataset consists of a total of 65536 triangular cells at a resolution of $513x65$.

**Red Sea** The third dataset is the Red Sea ensemble dataset from the 2020 VIS SciVis Contest [2,9,10]. The dataset shows simulated ocean currents and eddies in the Red Sea. These so-called eddies are vortices in ocean currents and are responsible, among other things, for the transport of salt in the ocean. Here, the dataset includes a time-dependent three-dimensional current field with corresponding velocity components $u$, $v$, and $w$, as well as sea surface satellite temperature information, sea level anomalies, and in situ salinity. These data are defined on a regular grid with a size of $499x499x49$ hexahedra cells. For this paper, we use the given mean dataset from the ensemble datasets and use only the sea surface (slice 0) from this dataset. This surface is also typically used for ocean research studies. This results in a dataset with 498002 triangular cells in the unstructured grid for our tests.

**Tensile Bar** As a final dataset, we use the Tensile Bar known from Tensorfield visualization [11,12]. Such tensile bars are used to investigate properties such as the strength of a material under tensile or compressive forces on a small sample. This typically involves introducing weak points such as notches or holes in the sample. The dataset used here was simulated using the commercial software Package Abaqus [1] and is a three-dimensional time-dependent dataset with 24 time steps into which a hole has been inserted. For our investigations, we use the time step 23 and a single layer in the middle of the data set at the thickness of 1.5. Since in this dataset, only one tensile force is applied in one direction and the dataset has only a small height of 4 layers, the use of a single layer is a suitable method to study the dataset. This results in a dataset with an unstructured grid of 9000 triangular cells on which a stress tensor is given at each point.

### 2.2 Recursion Strategies based on the Theoretical Case

As described in the main paper, our algorithm is based on joint recursion into two BVHs to generate a preselection of grid cells and control polygon edge pairs. For this purpose, we investigate which of the presented recursion strategies shown in Table 1 keeps the search effort the lowest on average compared to the acceleration structures of Klacansky et al. in which the search space is reduced by one BVH.

We use the theoretical case for our recursion strategy tests. In the theoretical case, we use all four datasets for our tests and combine them with the nine control polygons. The codomains for the four datasets are shown in Fig. 2a,b,c,d from the main paper. For the case (I), the number of intersection tests for all recursion strategies yields the results in Fig. 1. In this figure, the ratio to the number of intersection tests in Klacansky's method is plotted on a logarithmic scale. It shows that the results vary strongly between the individual recursion strategies, yet that all show an improvement on average.

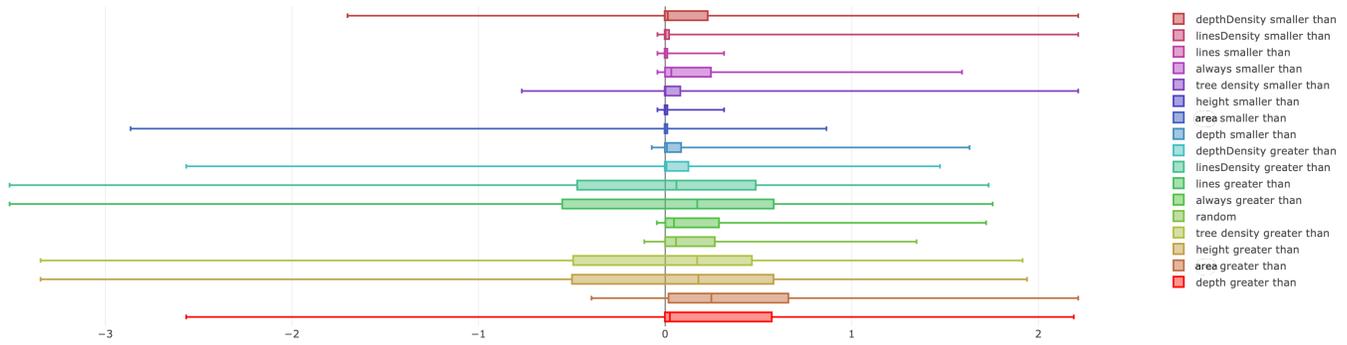

Figure 1: Plot to the Recursion Strategies.

Table 1: Overview of the Recursion Strategies

| | Strategies | |
|---|---|---|
| depth | greater than | smaller than |
| area | greater than | smaller than |
| tree height | greater than | smaller than |
| height | greater than | smaller than |
| tree density | greater than | smaller than |
| always | Start with greatest BVH | Start with smallest BVH |
| depthArea | greater than | smaller than |
| lines | greater than | smaller than |
| linesDensity | greater than | smaller than |
| depthDensity | greater than | smaller than |
| random | – | – |

Table 2: The effect of the number of elements per BVH leaf on the times of different algorithm parts. Times are in ms and average over multiple test cases. One can see that average number of intersection tests (NiT) decreases with increasing number of elements per BVH leaf, but the ratio of true positives vs all positives (TPaP) decreases and hence the extraction time increases. The best trade-off is 1 element per BVH leaf.

| Elements per Leaf | Traversal Time | Extra. Time | Total Time | NiT | TPaP |
|---|---|---|---|---|---|
| 16 | 0.30 | 4.04 | 4.44 | 5097.34 | 22.70% |
| 8 | 0.27 | 2.52 | 2.86 | 6727.74 | 29.10% |
| 4 | 0.29 | 1.70 | 2.08 | 9702.01 | 34.71% |
| 2 | 0.33 | 1.20 | 1.63 | 14393.72 | 45.00% |
| 1 | 0.40 | 0.86 | 1.37 | 21217.69 | 60.58% |

The best-performing recursion strategy is the *area greater than*. This strategy always subdivides the BVH that has the larger AABB. On average, the number of intersections of this strategy is 0.3 orders of magnitude smaller than Klacansky's and in the best case even more than 2 orders of magnitude smaller. For our further investigations, we, therefore, choose this recursion strategy.

### 2.3 Elements per Leaf based on the Theoretical Case (I)

The false positive rate has a great impact on the total time since the time for extraction is much more higher than the time for the acceleration structure. Therefore, the false positive rate should be reduced as much as possible. The maximum primitives per leaf affects this, so we test the sizes from the Table 2 in case (I). We use all the datasets and intersect them with the control polygons from Fig. 3 in the main paper. We position the center of the control polygons at 205 random positions in the range of AABB of the corresponding dataset codomains given in Fig. 2a,b,c,d from the main paper. Table 2 shows how the speed and search effort change across all 184,500 test cases. It shows how strongly the total time depends on the extraction time, which increases sharply with the magnitude of the false positive rate. In contrast, the contribution of the acceleration structure time to the total time is almost negligible, despite the fact that it increases with the search effort. Based on this investigation, we use 1 nodes per leaf for all further tests.

### 2.4 Example Visualization for Case (I) and Case (II)

In this section you can see example visualizations of the test cases in case (I) in Fig. 2 and case (II) in Fig. 3.

### 2.5 Example Visualization for Case (II)

This section visualizes the relationship between the total time and the number of control polygon edges for the three remaining datasets, Double Gyre in Fig. 5, Kármán vortex street in Fig. 4, and Tensile Bar in Fig. 6.

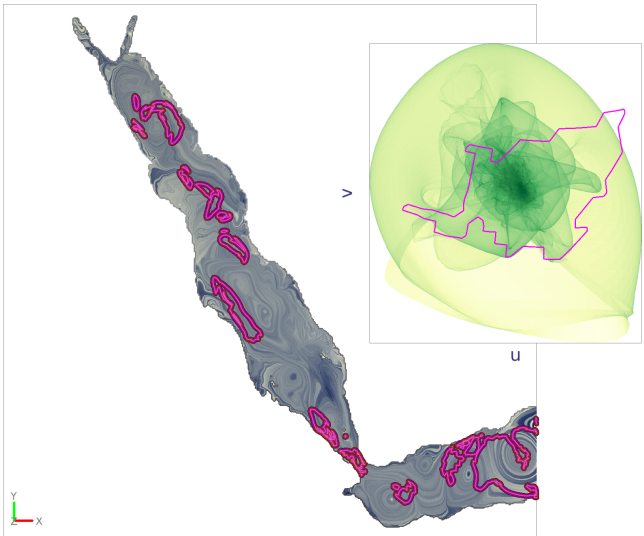

Figure 2: An example of the visualization of the tests in case (I) for the Red Sea dataset with the codomain from the main paper from Fig. 2c and the control polygon from the main paper from Fig. 3g with 126 edges.

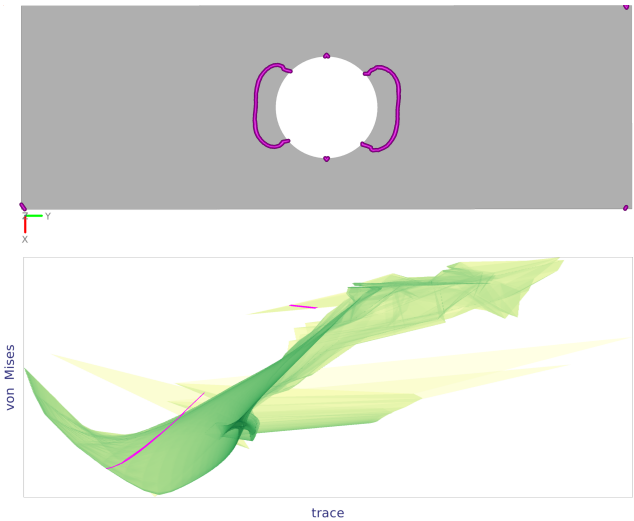

Figure 3: An example of the visualization of our experiments in case (II) for the Tensile Bar dataset with the codomain from Fig. 2h and an automatically generated control polygon with 120 edges.

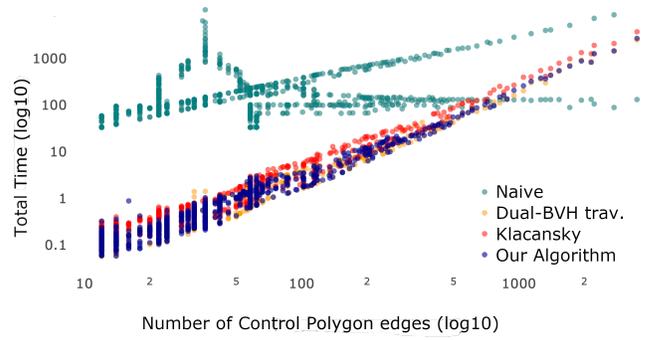

Figure 4: Plot the dataset Kármán vortex street in the case (II).

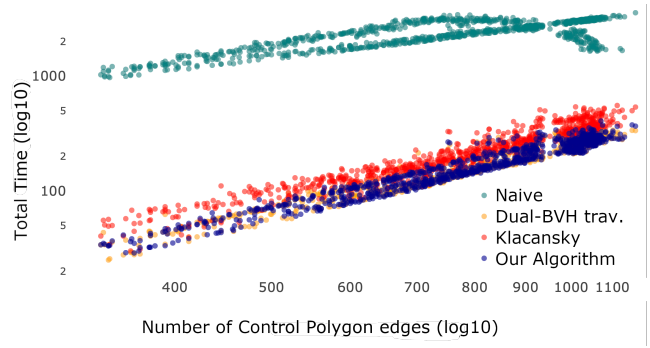

Figure 5: Plot the dataset Double Gyre in the case (II).

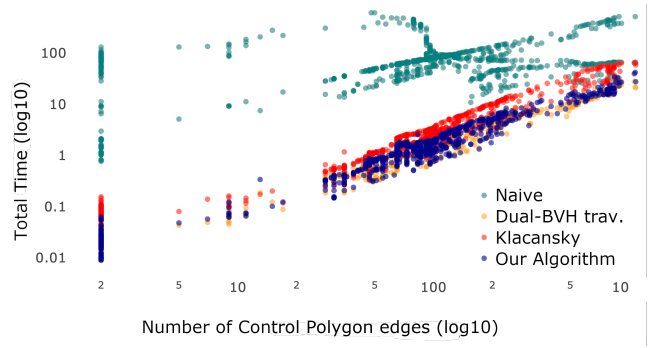

Figure 6: Plot the dataset Tensile Bar in the case (II).



\end{document}